\documentclass[twocolumn]{emulateapj}
\usepackage{graphicx}
\begin{document}

\newcommand{\ftd}{F_{\rm td}}
\newcommand{\ratio}{(R_{\rm app}/D)^2}
\newcommand{\fortyeight}{SAX~J1748.9$-$2021}
\newcommand{\twenty}{4U~1820$-$30}
\newcommand{\fortyfive}{EXO~1745$-$248}
\newcommand{\thirtyone}{KS~1731$-$260}
\newcommand{\oeight}{4U~1608$-$52}

\title{Statistics of Measuring Neutron Star Radii: Assessing A Frequentist and A Bayesian Approach}

\author{Feryal \"Ozel and Dimitrios Psaltis}

\affil{Departments of Astronomy and Physics, University of Arizona, 933 N. 
Cherry Ave., Tucson, AZ 85721, USA}

\begin{abstract}
Measuring neutron star radii with spectroscopic and timing techniques
relies on the combination of multiple observables to break the
degeneracies between the mass and radius introduced by general
relativistic effects. Here, we explore a previously used frequentist
and a newly proposed Bayesian framework to obtain the most likely
value and the uncertainty in such a measurement. We find that, for the
expected range of masses and radii and for realistic measurement
errors, the frequentist approach suffers from biases that are larger
than the accuracy in the radius measurement required to distinguish
between the different equations of state. In contrast, in the Bayesian
framework, the inferred uncertainties are larger, but the most likely
values do not suffer from such biases. We also investigate ways of
quantifying the degree of consistency between different spectroscopic
measurements from a single source. We show that a careful assessment
of the systematic uncertainties in the measurements eliminates the
need for introducing ad hoc biases, which lead to artificially large
inferred radii.
\end{abstract}


\section{INTRODUCTION}
There has been significant recent interest in measuring the radii of
neutron stars. The radii have been shown to be direct probes of the
ultradense matter equation of state, which, in turn is connected to
numerous astrophysical phenomena, such as the dynamics and outcomes of
stellar explosions and the signals from the coalescense of compact
objects. 

Because of strong general relativistic effects, all observables from
the vicinity of a neutron star depend on different combinations of its
mass and radius. (For fast spinning neutron stars, higher order
moments such as the spin and quadrupole also play a role). As a
result, measuring radii require at least two distinct observables to
break the degeneracy with mass. 

During the last decade, measurements of neutron star radii have
primarily relied on utilizing two spectroscopic observables from
neutron stars that show thermonuclear bursts: the apparent angular
size and the Eddington flux (\"Ozel 2006). This technique has been
applied to a number of neutron stars (see., e.g., \"Ozel et al.\ 2009;
G\"uver et al.\ 2010a, b) and led to radius measurements that
significantly constrained the ultradense matter equation of state
(\"Ozel et al. 2010). Similar spectroscopic techniques have also been
explored that rely instead on the apparent angular size and the
Eddington temperature (Suleimanov et al.\ 2012) or the apparent
angular size and the evolution of the spectral temperature (e.g.,
Kusmierek et al.\ 2011). Measurements in the near future with NICER
(Gendreau et al.\ 2012) and LOFT (Feroci et al.\ 2012) will utilize
different properties of the pulse profiles observed from the surfaces
of spinning neutron stars. In this case, the harmonic content of the
pulse profiles and their energy dependence provide the necessary
observables that help break the degeneracy between the neutron-star
radius and mass (Psaltis et al.\ 2014).

In all of these approaches, the measurement of two observables leads,
in principle, to a solution for the two unknowns of interest, namely
the neutron-star radius and mass. In practice, however, the situation
is more complicated because of the nonlinear dependence of the
observables on the neutron-star parameters. Depending on their
particular values, a set of measurements may lead to two, one, or no
solutions for the mass and radius. Because of these nonlinearities,
the inferences from such measurements likely depend on the particular
statistical estimators used in combining these constraints.

In this paper, we assess, using mock data, the previously used
frequentist method for measuring neutron star radii based on two
spectroscopic observables. We show that the radii inferred with this
frequentist approach are often biased and, in some cases, the true
solution is formally excluded. The level of bias increases
significantly as the measurement errors increase. We then devise a
approach within a Bayesian framework that is substantially less
affected by these nonlinearities, even in the case of large
uncertainties. We also discuss the conditions under which the lack of
solutions for a large fraction of the parameter space can be used to
infer the presence of systematic uncertainties in the measurements.

\section{THE PREVIOUS APPROACH TO DETERMINING RADII}

We will focus hereafter on the measurement of the neutron star radii
and masses based on the combination of the apparent angular size and
the Eddington flux observed during thermonuclear bursts. Even though
we use this as our primary example, the results are general and can be
translated to the other combinations of observables discussed in the
introduction. 

We follow the formalism in \"Ozel et al.\ (2009) and assume that the
apparent angular size $A$ and the Eddington flux $F_{\rm td}$ have
been measured during the cooling tails and the touchdown moments of
thermonuclear bursts, respectively. In the Schwarzschild approximation, 
which is appropriate for slowly spinning neutron stars, these quantities 
are related to the neutron star mass $M$ and radius $R$ according to 
\begin{equation}
A = \frac{R^2}{D^2 f_c^4}\left(1-\frac{2 G M}{R c^2}\right)^{-1}
\label{eq:a}
\end{equation}
and 
\begin{equation}
F_{\rm td} = \frac{GMc}{k_{\rm es} D^2}\left(1-\frac{2 G M}{R c^2}\right)^{1/2},
\label{eq:ftd}
\end{equation}
where $G$ is the gravitational constant, $c$ is the speed of light,
$D$ is the distance to the neutron star, $f_c$ is the color correction
factor due to the stellar atmosphere, $k_{\rm es}=0.2 \;
(1+X)$~cm$^2$~g$^{-1}$ is the electron scattering opacity, and $X$ is
the hydrogen mass fraction. Note that for moderately spinning stars,
there are additional corrections that depend on the spin and the
quadrupole moment of the neutron star (see Baub\"ock et al.\ 2015). In
addition, the Eddington flux is subject to temperature-dependent
corrections due to the energy-dependent terms in the Klein-Nishina
cross section, as discussed in Paczynski (1983). For simplicity, we
ignore these corrections in the present statistical treatment; see,
however, \"Ozel et al. (2015) for their effects on the inferred radii
and masses.

In principle, these two equations can be solved for the neutron star
mass and radius given the observables. Because of the nonlinear nature
of these equations, there can be zero, one, or two solutions. The
number of solutions depends on the value of the quantity (see, e.g.,
Steiner et al.\ 2010)
\begin{equation}
\alpha = \frac{F_{\rm td} k_{\rm es} D}{c^3 f_c^2 A^{1/2}}
\label{eq:alpha}
\end{equation}
such that, when $\alpha >1/8$, there are no solutions, when $\alpha =
1/8$, there is one (double) solution, and when $\alpha < 1/8$, there
are two distinct solutions. The critical value occurs when $R=4GM/c^2$
independent of all the other parameters. 

When this method is applied in practice, the uncertainties inherent in
the measurements of $A$ and $F_{\rm td}$ need to be converted into
uncertainties in $R$ and $M$. In this frequentist approach, this is
achieved by sampling the likelihoods over the observables, calculating
the mass and radius for each pair of observables, and using this to
populate a posterior likelihood over radius and mass. In \"Ozel et al.
(2009), this was carried out analytically using the Jacobian
transformations of the posterior likelihoods. Steiner et al.\ (2010)
repeated the same frequentist analysis using Monte Carlo techniques to
sample the likelihoods over the observables. Both methods of
calculation give identical results under the same set of assumptions.

In this formalism, the posterior likelihood over mass and radius is 
given by
\begin{eqnarray}
&& P(M,R) dM dR = \frac{1}{2} \int P(D) \; dD \int P(f_c) \; df_c \int P(X) 
\; dX \nonumber \\ && \times P[F_{\rm td}(M, R, D)] 
P[A(M, R, D)] \; J\left(\frac{\ftd,A}{M,R}\right) dMdR.
\label{eq:likelihood}
\end{eqnarray}
Here, $P(D)$, $P(\ftd)$, and $P(A)$ are the likelihoods over the
distance, touchdown flux, and the apparent angular size measurements,
respectively.  $P(X)$ and $P(f_c)$ are the priors over the hydrogen
mass fraction and the color correction factor. Finally,
\begin{equation}
J\left(\frac{\ftd,A}{M,R}\right) = \frac{2 c G R \left\vert1-\frac{4 G M}{R c^2}\right\vert}
{D^4 f_c^4 k_{\rm{es}} \left(1-\frac{2 G M}{R c^2}\right)^{3/2}}
\label{eq:jac}
\end{equation}
is the Jacobian of the transformation. The factor $1/2$ appears in
this equation because nearly all pairs of observables correspond to
two distinct pairs of $M$ and $R$. The region of the parameter space
for which the observables correspond to one repeating solution for $M$
and $R$ has zero volume and will, therefore, not contribute to the
final likelihood.

It is evident from equations~(\ref{eq:likelihood}) and (\ref{eq:jac})
that $P(M,R)$ is identically equal to zero when $R=4GM/c^2$, independent
of the measurements. This occurs because for masses and radii that
satisfy this condition, the two constraints imposed by the observables
(eqs. \ref{eq:a} and \ref{eq:ftd}) are parallel at the point of
contact and the system is degenerate. As a result, if the mass and
radius of the neutron star lie along the $R=4GM/c^2$ line, which is
very likely for the expected range of neutron star masses and radii,
the posterior likelihood will exclude the true solution and, hence,
will introduce a bias in this case. 

\begin{figure*}
\centering
   \includegraphics[scale=0.4]{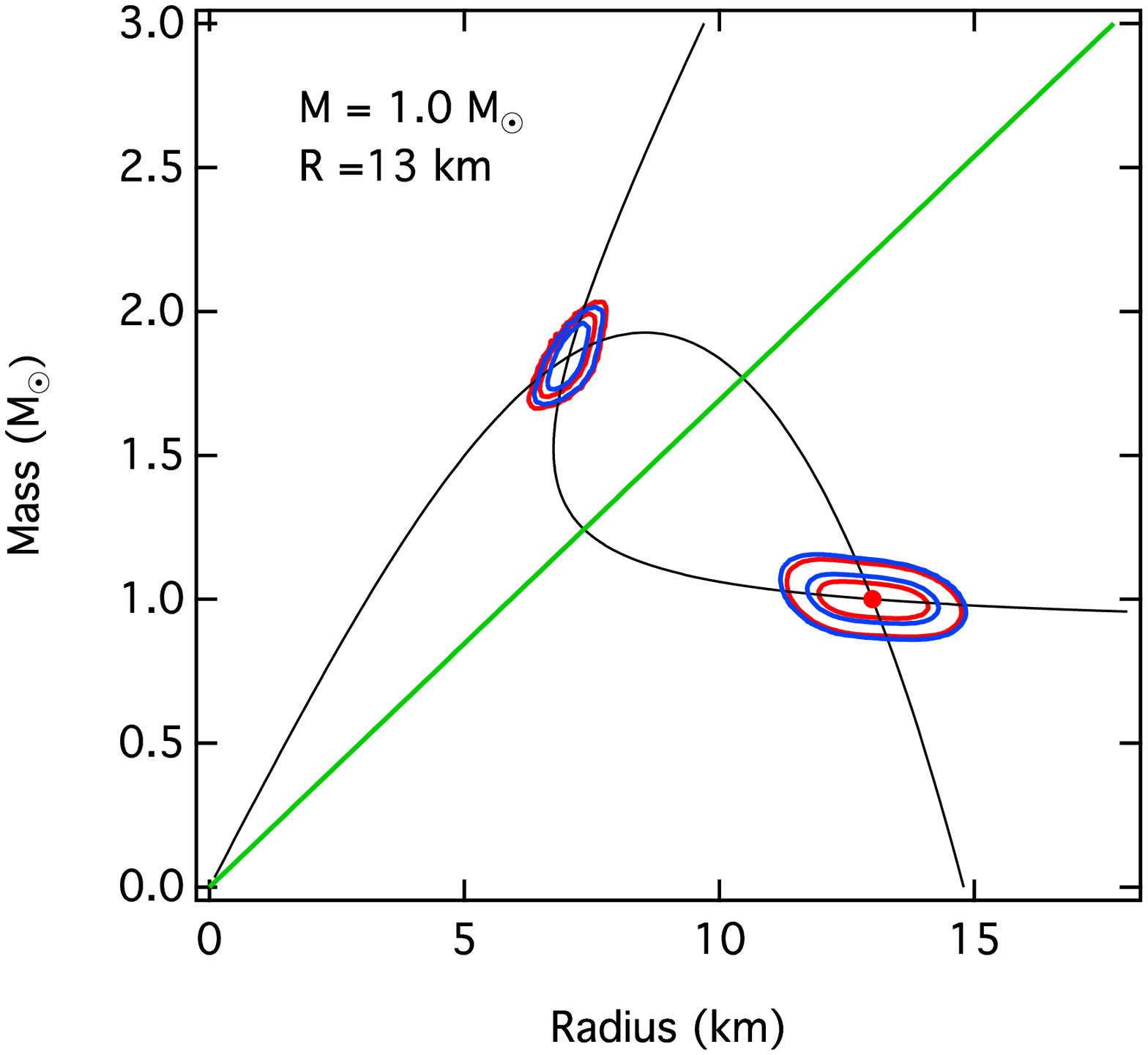}
   \includegraphics[scale=0.4]{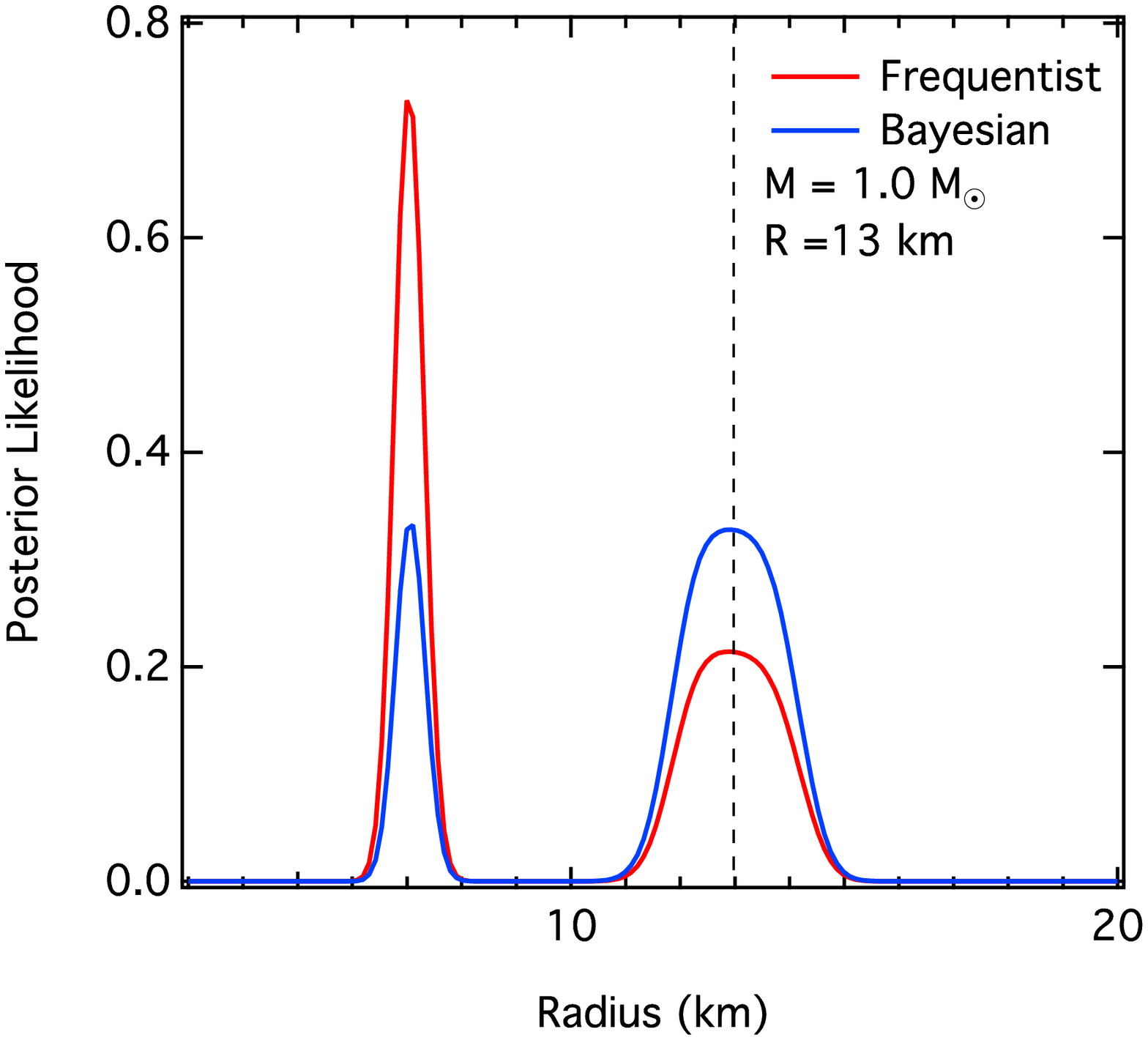}
   \includegraphics[scale=0.4]{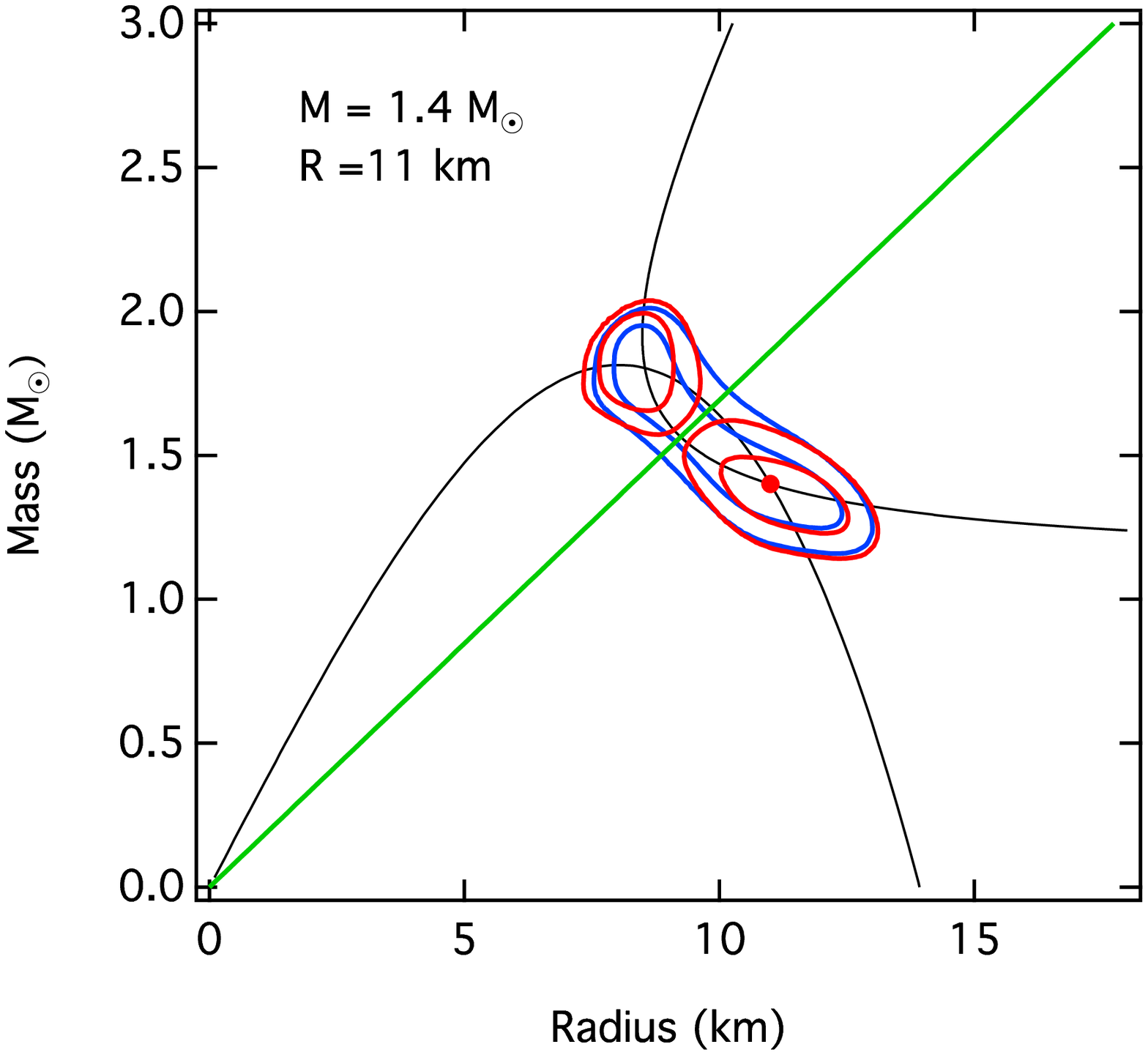}
   \includegraphics[scale=0.4]{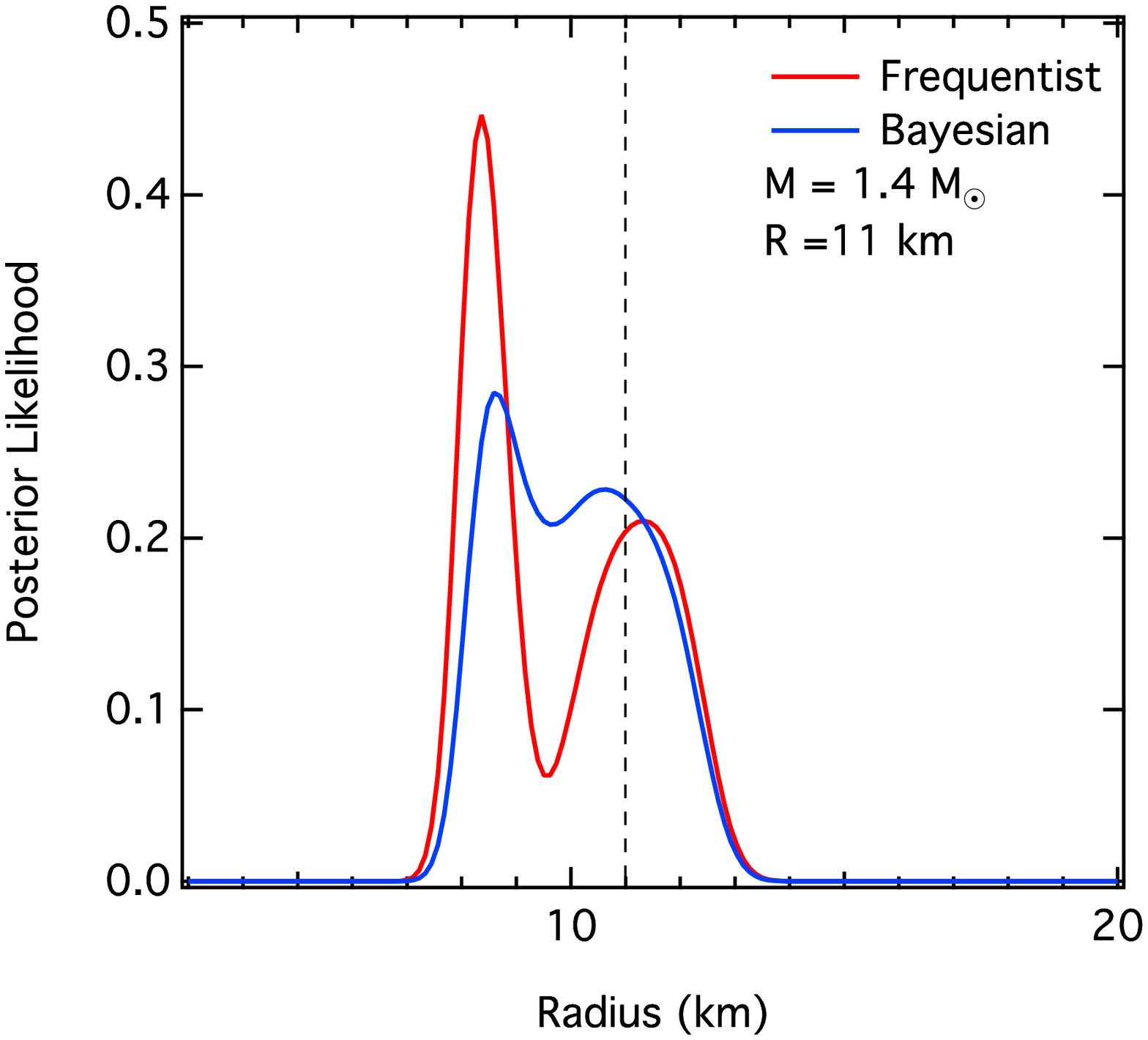}
   \includegraphics[scale=0.4]{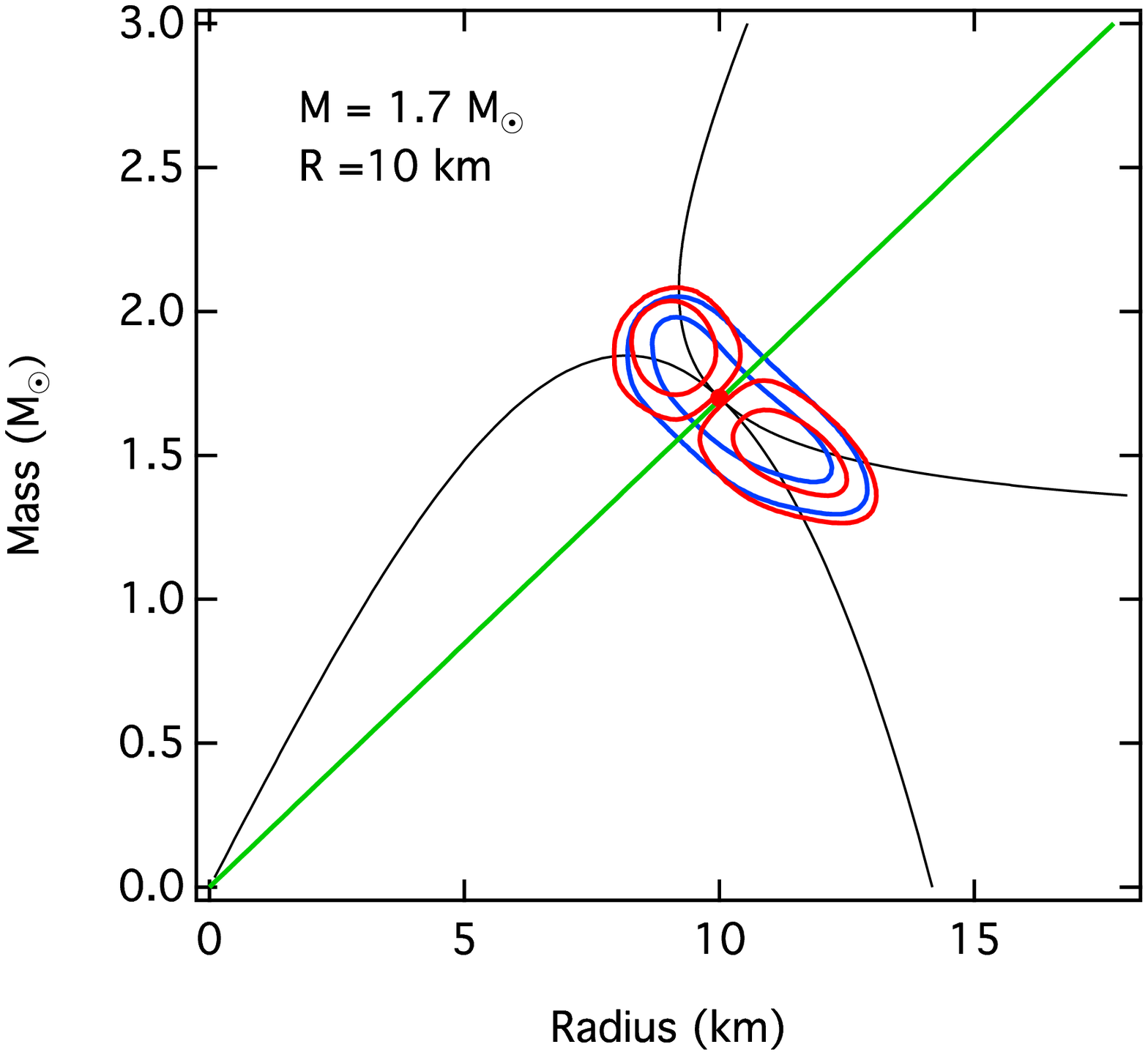}
   \includegraphics[scale=0.4]{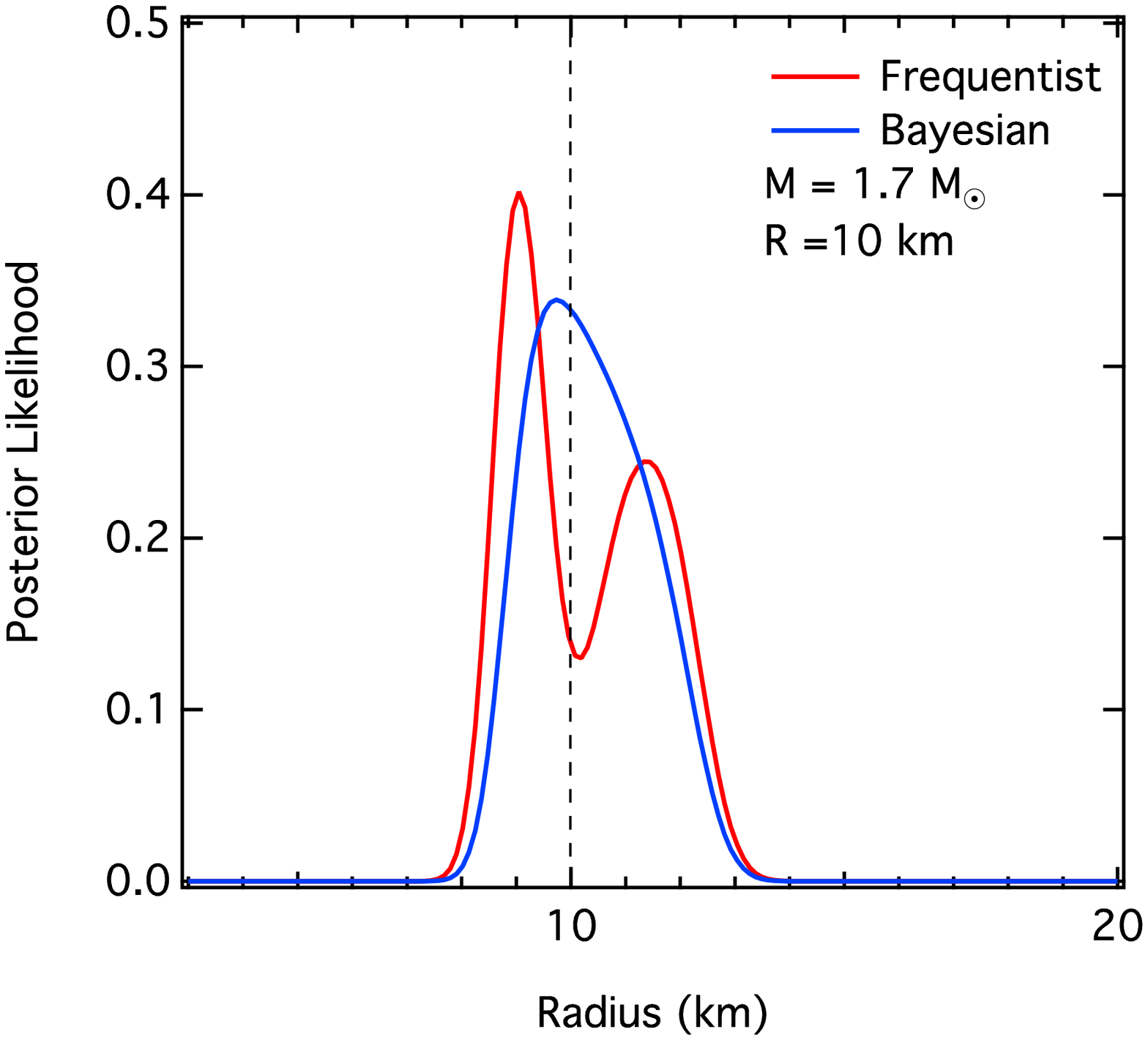}
\caption{{\em (Left)\/} Contours of 68\% and 95\% posterior
  likelihoods over mass and radius using {\em (red)\/} the frequentist
  and {\em (blue)\/} Bayesian approaches discussed in the text, for
  three neutron stars with different assumed masses and radii. The
  true masses and radii are denoted by red filled circles. The black
  curves correspond to lines of constant apparent angular size, $A$,
  and Eddington flux, $\ftd$. The green line corresponds to the
  critical curve $R=4GM/c^2$. {\em (Right)\/} The same posterior
  likelihoods marginalized over mass. As the true mass and radius of a
  neutron star approaches the critical curve, the frequentist approach
  significantly biases the inferred radii. In all cases, the distance
  of the neutron star was fixed to 4~kpc, the color correction factor
  to $f_c=1.35$, and the hydrogen mass fraction to $X=0$.}
\label{fig1} 
\end{figure*}

To explore the severity of this bias for a range of neutron star
masses and radii, we perform the frequentist inference described above
for three different pairs of mock measurements that correspond to
stars with different $R/M$ ratios. In all cases, we calculate the
apparent angular size and the Eddington flux for the assumed mass and
radius of the neutron star by fixing the hydrogen mass fraction to
$X=0$, the distance to 4~kpc, and the color correction factor to 1.35.
We then assign a 5\% Gaussian error in $A$ and $\ftd$. To infer $R$
and $M$ from these mock measurements, we assume a perfect prior
knowledge of the hydrogen mass fraction and distance but take a boxcar
prior in the color correction factor between 1.3 and 1.4. The left
panels of Figure~\ref{fig1} show the 68\% and 95\% confidence contours
of the posterior likelihood over the inferred mass and radius while
the right panels show the posterior likelihood over radius when
$P(R,M)$ is marginalized over mass. In the left panels, the black
lines correspond to contours of constant $A$ and $\ftd$ at their
central values and for $f_c=1.35$, such that one of the intersection
of these two lines correspond to the assumed radius.  The green line
is the critical curve $R=4GM/c^2$.

The examples depicted in Figure~1 show that when the assumed value of
$R/M$ is significantly away from the critical curve, the posterior
likelihood of this frequentist approach is centered on the true
solution, with little evidence for bias. However, as the assumed ratio
approaches the critical curve, which also corresponds to more
realistic values of masses and radii, the inferred radii can be
substantially biased toward higher or lower values, by as much as
1.5~km. This is an unacceptably large bias given that $\lesssim 1$~km
precision is required in the radius measurements in order to place
meaningful constraints on the equation of state (see, e.g., \"Ozel \&
Psaltis 2009). Moreover, in this idealized example, we assumed that
the distance is known with extremely high accuracy, which is
unrealistic for the neutron star sources used for the radius
measurements. In Section 4, we will explore the effect of realistic
distance uncertainties on the radius measurement bias and show that it
gets even worse in this frequentist approach. For the measurements
performed to date using this approach, this implies that the inferred
radii are likely to have been overestimated by $\sim 1-1.5$~km.

\section{A NEW BAYESIAN APPROACH TO DETERMINING RADII}

The posterior likelihoods over the neutron star mass and radius 
can be inferred from the same observables $P(A)$ and $P(\ftd)$ 
in a different way using Bayes' theorem, according to which 
\begin{equation}
P(M,R |{\rm data}) = C P(\rm{data} | M, R) P_{\rm pri}(M) P_{\rm pri} (R),  
\end{equation}
where $P_{\rm pri} (M)$ and $P_{\rm pri} (R)$ are the priors over the
mass and radius and $C$ is an appropriate normalization constant. 
Given that $A$ and $\ftd$ are ideally uncorrelated measurements, we
can write 
\begin{eqnarray}
P(\rm{data} | M, R) &&= \int P(D) \; dD \int P(f_c) \; df_c \int P(X) 
\; dX \nonumber \\ && \times P[F_{\rm td}(M, R, D)] 
P[A(M, R, D)].
\label{eq:bayes}
\end{eqnarray}
Assuming a flat prior over the radius and mass, it is clear from
equations~(\ref{eq:likelihood}) and (\ref{eq:bayes}) that the
difference between the Bayesian and frequentist inferences discussed
here is the presence of the Jacobian in the latter. Consequently, the
biases introduced in the frequentist approach by the zero points of
the Jacobian are not present here. The underlying reason is that in a
Bayesian framework, one does not ask what is the mass and radius pair
that corresponds to each set of observables, but rather, for a given
mass and radius, what is the chance of obtaining the corresponding set
of observables.

In the left panels of Figure~\ref{fig1}, we overplot the 68\% and 95\%
confidence contours of the posterior likelihood calculated according
to equation~\ref{eq:bayes} and show the marginalized likelihood over
the radius in the right panels. When the $R/M$ ratio is significantly
away from the critical curve, the Bayesian approach gives unbiased
results as in the frequentist approach. There is still a difference
between the two approaches, however, which becomes evident in the
marginalized likelihoods: the Bayesian approach gives equal maximum
likelihood between the two equivalent solutions, whereas the
frequentist yields equal integrated likelihoods in the two islands.
Given that the two solutions are mathematically equivalent (albeit the
smaller radius is sometimes physically unacceptable), the difference
in the integrated posterior likelihoods between the two islands should
not be used to discriminate between the two solutions.

The biggest difference in the two frameworks occurs when the $R/M$
ratio of neutron stars approach or lie on the critical curve. In these
cases, the 68\% confidence contour of the Bayesian posterior
likelihood encompasses both solutions and, both in the full $P(M, R)$
and in the marginalized $P(R)$ likelihoods, the method results in
larger uncertainties in the inference of the radii than in the
frequentist approach. However, it does not suffer from the biases of
the frequentist approach, which are exacerbated when the neutron stars
lie on the critical curve (see the bottom panels of Fig.~\ref{fig1}).

\section{BIASES IN THE INFERRED EQUATION OF STATE FROM MULTIPLE RADIUS MEASUREMENTS}

In the previous sections, we addressed the biases in the inferred
radius of a single star using two approaches, a previously developed
frequentist and a newly proposed Bayesian formalism, under the
assumption of small uncertainties in the two measured quantities $A$
and $\ftd$. In reality, multiple radius measurements, with different
sources of uncertainty, are necessary to place significant constraints
on the dense matter equation of state. In this section, we compare the
performances of these two approaches under frequentist
considerations. In particular, we explore the uncertainties and biases
in the inferred properties of the equation of state using mock data
sets for multiple objects.

Earlier work has shown that all equations of state that smoothly
connect to the low density equation of state of matter can be
represented by a small number of parameters, which can be inferred by
the measurement of several neutron star radii at different masses
(Lindblom 1992; Read et al.\ 2009; \"Ozel \& Psaltis 2009; Gandolfi et
al.\ 2014). In the relevant range of observed neutron-star masses
(see, e.g., \"Ozel et al.\ 2012), most equations of state predict
nearly constant radii. For simplicity in the present study, we make
use of this fact to reduce the complexity of the equation of state and
represent them by a single parameter, which we take to be the constant
radius.

\begin{figure}
\centering \includegraphics[scale=0.4]{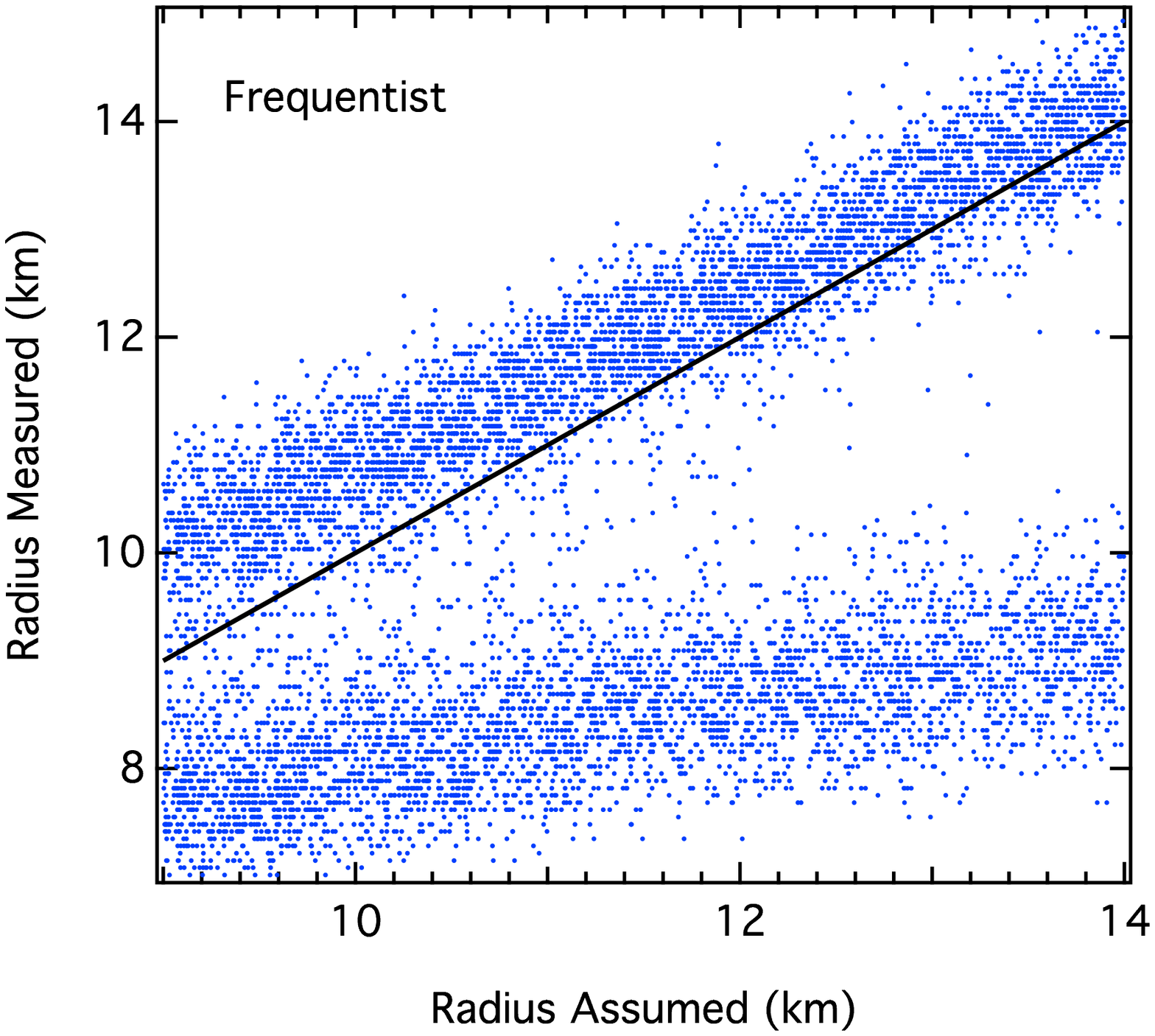}
\includegraphics[scale=0.4]{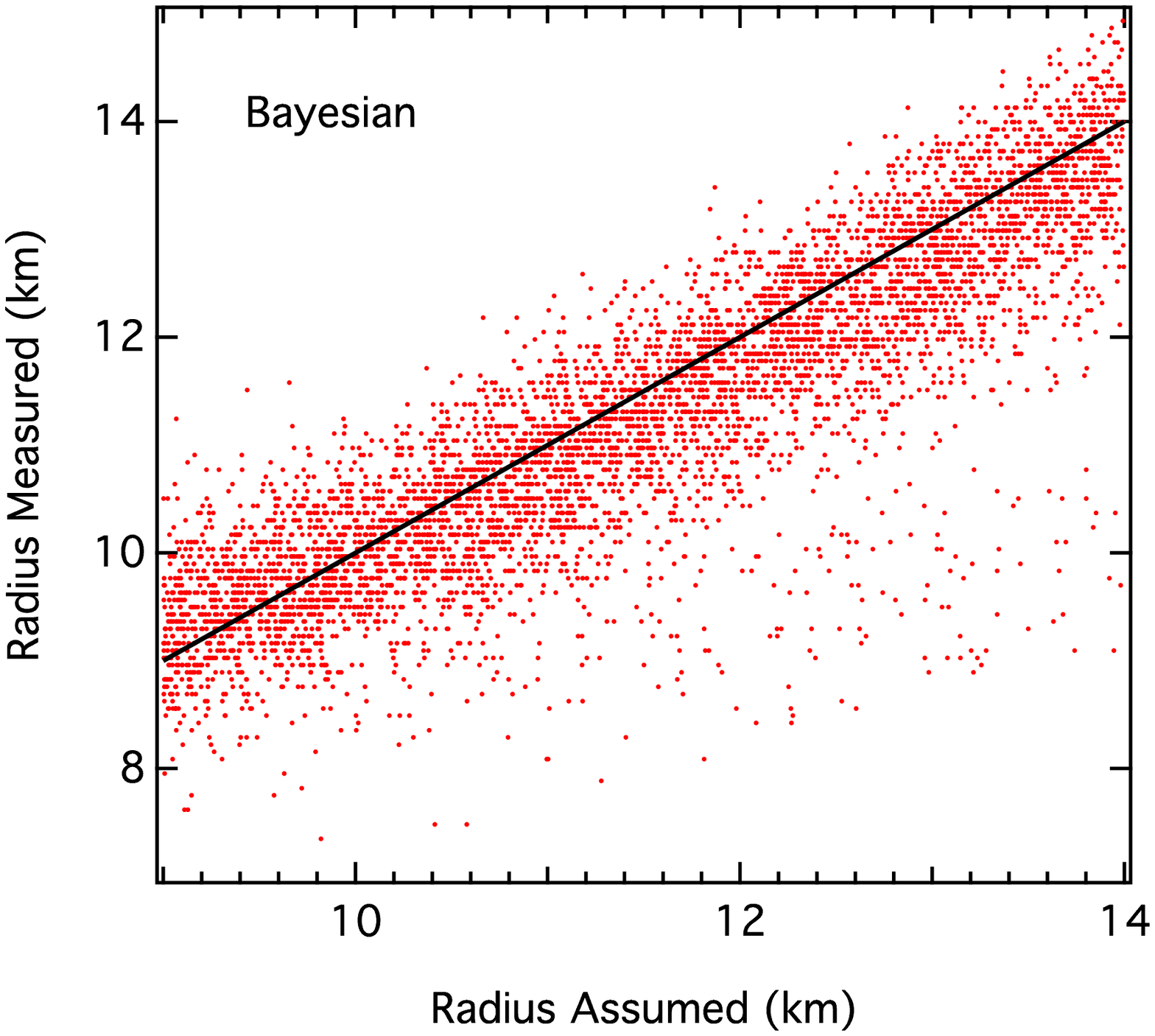}
\caption{The radius inferred for an equation of state (chosen here to
  predict constant radius neutron stars over the mass range of
  interest) using mock spectroscopic observations of five neutron
  stars, plotted against the assumed radius that it predicts, for a
  large ensemble of realizations of the mock data, in the {\em
    (upper)\/} frequentist and {\em (lower)\/} Bayesian approaches
  discussed in the text. All simulated neutron stars were placed at a
  distance of 4~kpc and their masses were drawn from a Gaussian
  distribution that has been inferred observationally for the
  decendendents of X-ray bursters. The spectroscopic measurements were
  assumed to have 10\% Gaussian errors, while the distance and
  hydrogen mass fraction was assumed to be accurately known a priori.
  The frequentist approach leads to substantial biases in the infered
  radii, while the Bayesian approach does not suffer from biases.}
\label{fig2}
\end{figure}

For each assumed neutron star equation of state, and, hence, assumed
radius, we draw five neutron star masses from a Gaussian distribution
with a mean of $1.46~M_\odot$ and a dispersion of $0.21~M_\odot$; this
is the distribution of masses that is observationally inferred for the
descendants of X-ray bursters (\"Ozel et al.\ 2012). For each of these
neutron stars, we calculate the Eddington flux and the apparent
angular size assuming a distance of 4~kpc, color correction factor of
1.35, and a hydrogen mass fraction $X=0$. We assign a 10\% Gaussian
uncertainty in each of these values, which is typical for X-ray
bursters (see G\"uver et al. 2012 a,b) and draw randomly a set of mock
measurements from these distributions. We then apply the frequentist
and Bayesian frameworks discussed above to each of these measurements,
assuming that the distance and the hydrogen mass fraction are known a
priori, but allow the color correction factor to be in the range
1.3-1.4. We marginalize the five resulting posterior likelihoods over
mass and then multiply them so that we obtain the posterior likelihood
over the inferred radius for that equation of state. We repeat this
procedure for 10,000 equations of state with assumed radii between 9
and 14~km and different realizations of the neutron star masses drawn
from the same distribution.

\begin{figure}
\centering
   \includegraphics[scale=0.4]{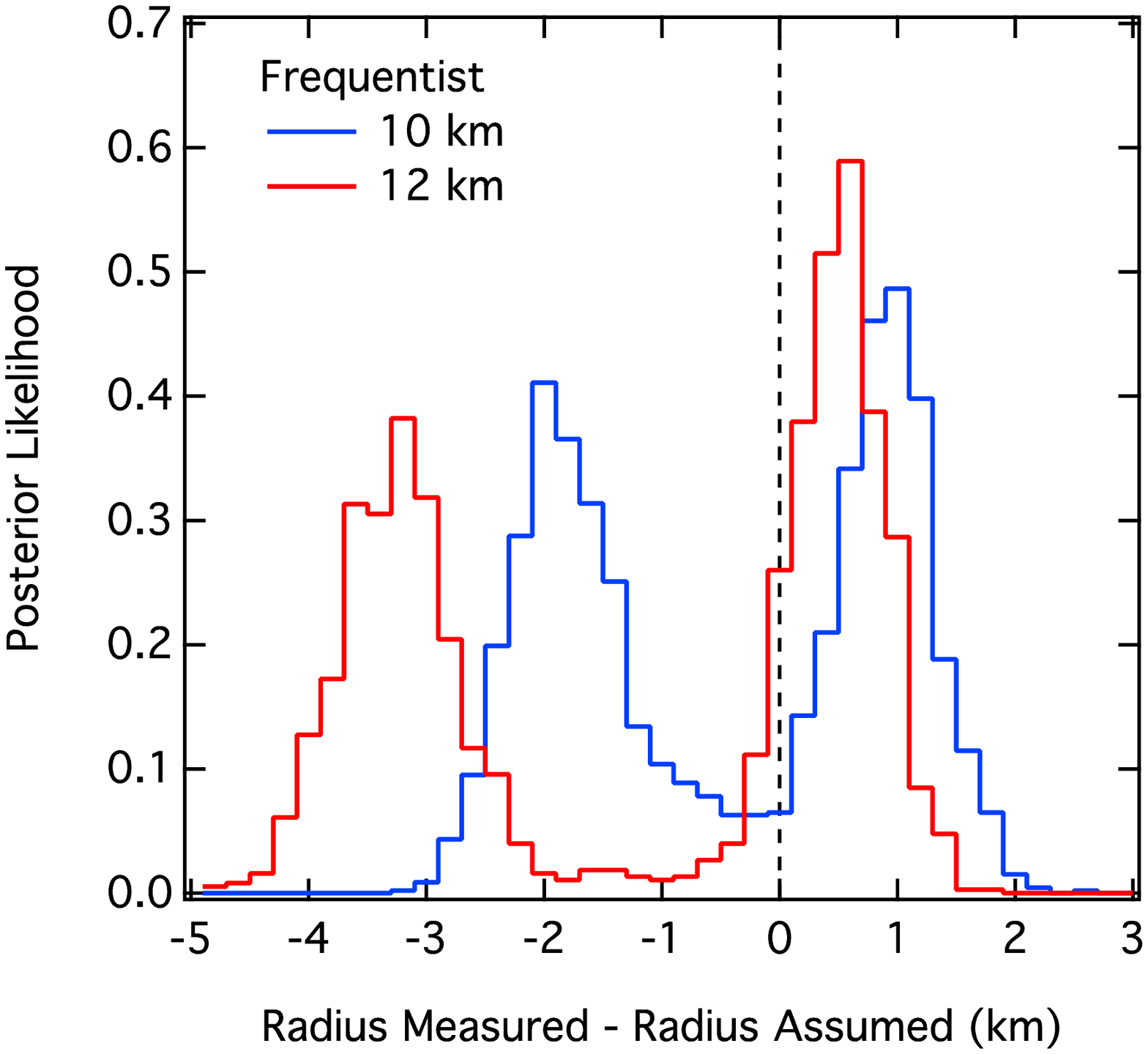}
   \includegraphics[scale=0.4]{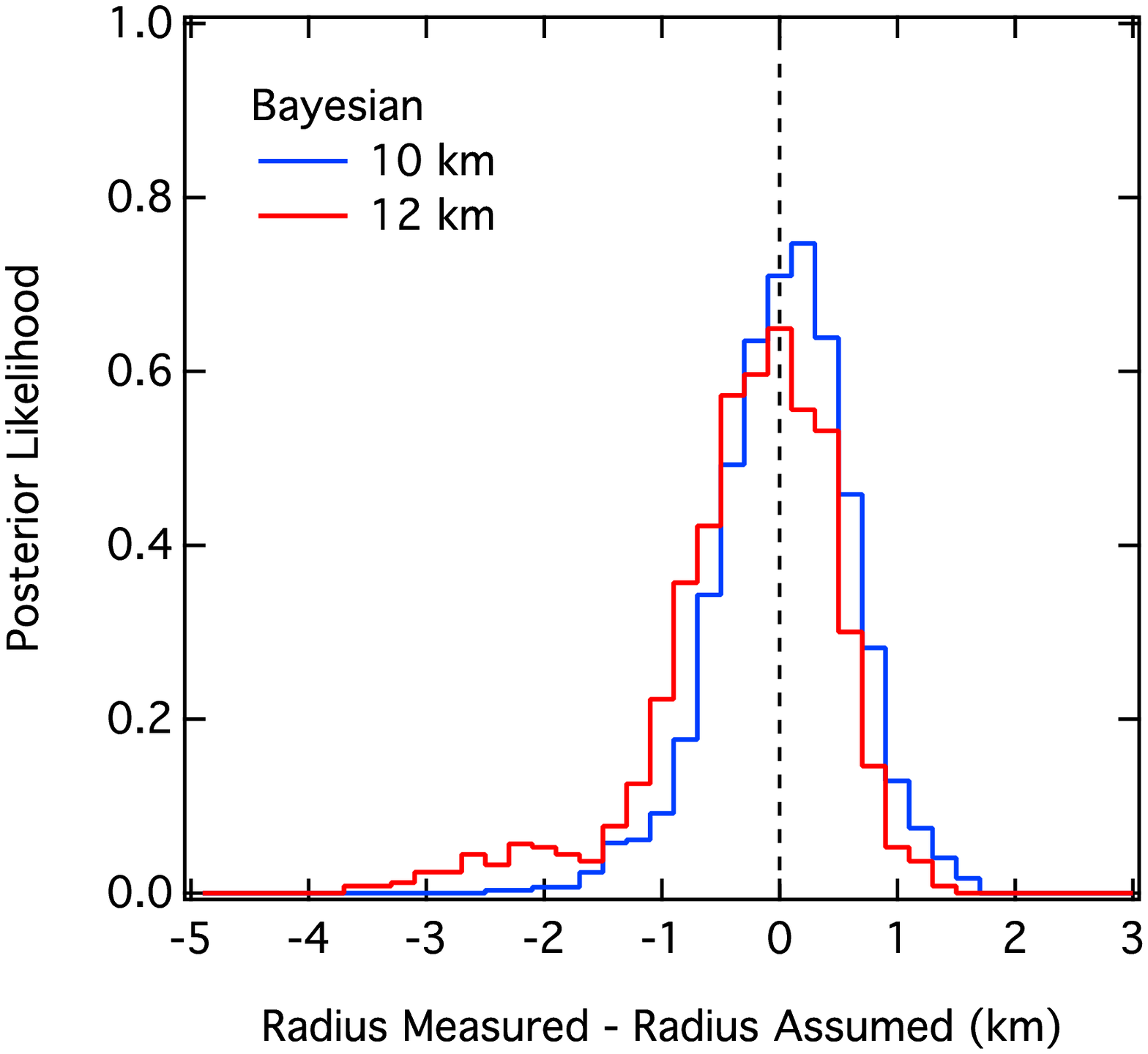}
\caption{The histogram of the difference between the inferred and
  assumed radii for two different values of the assumed radius, for
  the simulated data shown in Figure~\ref{fig2}. The bias toward
  larger radii in the frequentist approach is $\sim 0.75$~km for 12~km
  neutron stars and $\sim 1$~km for 10~km stars. The proposed Bayesian
  inference is more accurate but less precise.}
\label{fig3}
\end{figure}

In Figure~\ref{fig2}, we plot the inferred vs. the assumed radius for
each set of mock data for the frequentist and the Bayesian inferences.
As expected from the discussion in Section~2, the previously used
frequentist inference yields two well separated tracks of solutions,
whereas the proposed Bayesian approach has a single band. (Note that
the few outliers correspond to the unlikely situations where all five
drawn masses are comparable and place the neutron stars away from the
critical curve). More importantly, even if the band of small radii is
rejected on physical grounds, the remaining band in the frequentist
approach yields measured radii that are biased toward larger
values. In contrast, the radii inferred in the Bayesian formalism are
centered on the assumed values, with little, if any, hints of bias.

To quantify the degree of bias in these measurements, we plot in
Figure~\ref{fig3} the histogram of the difference between the inferred
and assumed radii for two different values of the assumed radius.  The
bias in the frequentist approach (again focusing on the larger of the
two solutions) is $+0.75$~km for a 12~km neutron star and $+1$~km for
a 10~km neutron star. The histogram in the Bayesian approach, on the
other hand, is centered on the true value for both assumed radii, even
though it is wider than either of the frequentist peaks. In other
words, the frequentist approach generates the more precise but less
accurate solution.

The last aspect of the measurements that may bias the inferred radii
is how well the distance to the source is known, as reflected in the
width of the prior over the distance $P(D)$ that is considered in the
calculation. This is of particular concern in the frequentist approach
because the Jacobian of the transformation scales as $D^4$ and,
therefore, places large weight on the smaller distances. We repeated
the above simulation of mock observations with two differences. First,
we reduced the assumed uncertainties in the measurements of $A$ and
$\ftd$ to 5\% in order to focus on the effect of the broader range of
distances. Second, even though we placed all simulated neutron stars
at 4~kpc, we assumed a flat prior in the distance in the range 3-5~kpc
when inferring the radii from the mock measurements.

\begin{figure}
\centering
   \includegraphics[scale=0.4]{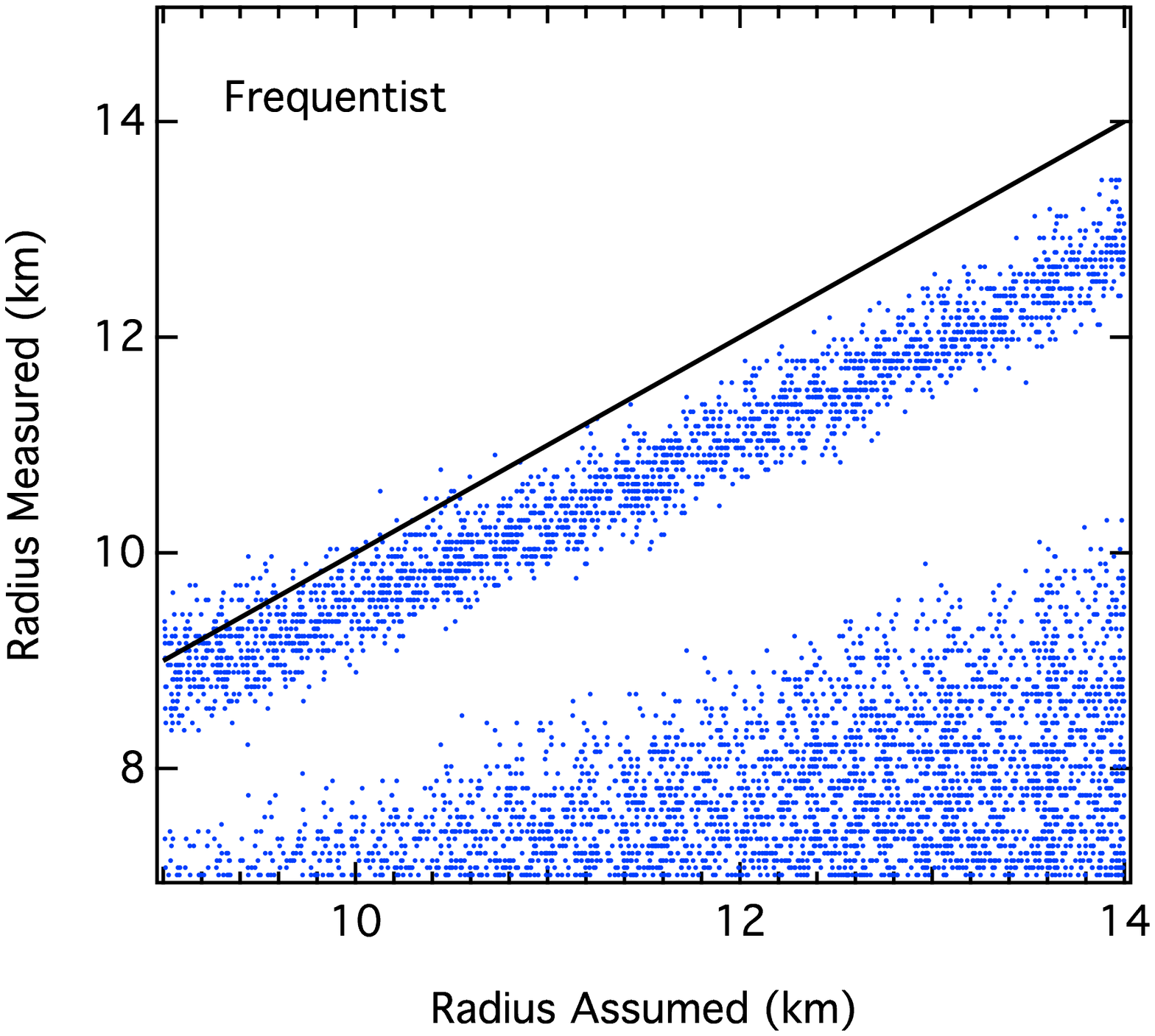}
   \includegraphics[scale=0.4]{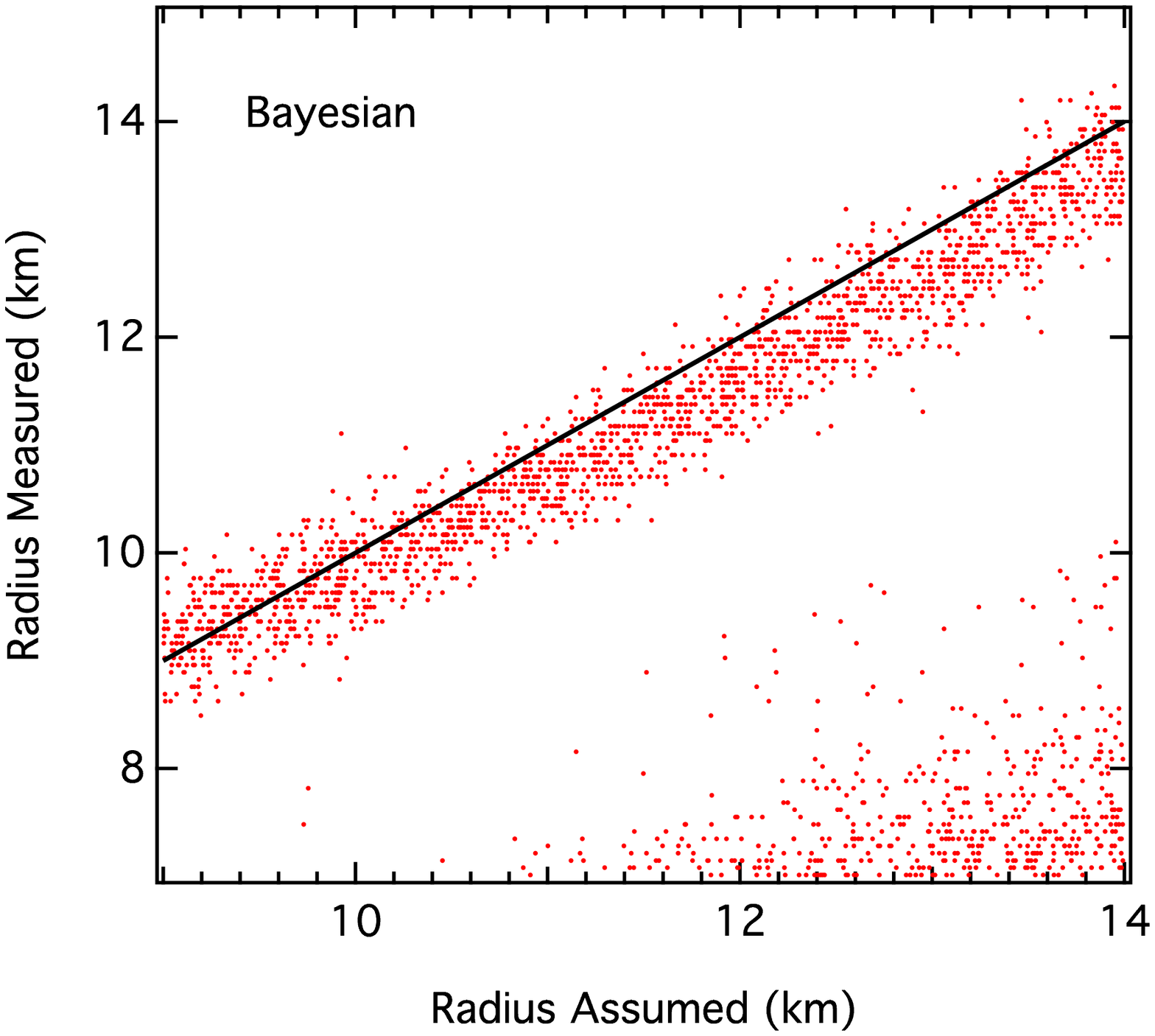}
\caption{Same as Figure~\ref{fig2} but for assumed 5\% uncertainties
  in the spectroscopic measurements and a flat prior in distances in
  the range 3-5~kpc. In this case, the frequentist approach shows a
  significant bias toward lower radii, while the Bayesian approach is
  only marginally affected.}
\label{fig4}
\end{figure}

In Figure~\ref{fig4}, we plot the inferred vs. the assumed radius for
each set of mock data when the range of distances in the priors
dominate the uncertainties of the measurement. As in the previous set
of mock data, there are two bands of acceptable solutions. However, we
will not consider the bands that correspond to small radii any
further, as these can typically be rejected on physical grounds. In
the frequentist inference, the large range of distance priors biases
the inferred radii toward values that are smaller by as much as 1~km
than the assumed radii. The reason for this bias is the strong
dependence of the Jacobian on distance, which favors the smaller
distances and hence the smaller radii. In contrast, the Bayesian
inference suffers from a much smaller bias toward smaller radii, even
in this case.

\section{THE POSTERIOR PROBABILITIES OF THE CONSISTENT MASS-RADIUS SOLUTIONS}

In the previous sections, we investigated the biases and uncertainties
in the spectroscopically inferred masses and radii of neutron stars
within a frequentist and a Bayesian framework. We showed that the fact
that real solutions for mass and radius can be obtained for sets of
observables that satify $\alpha \leq 1/8$ (see eq.~\ref{eq:alpha})
introduces significant biases in this frequentist approach, especially
when the true $M,R$ of the neutron star lies near this critical value.

\begin{figure}[t]
\centering
   \includegraphics[scale=0.4]{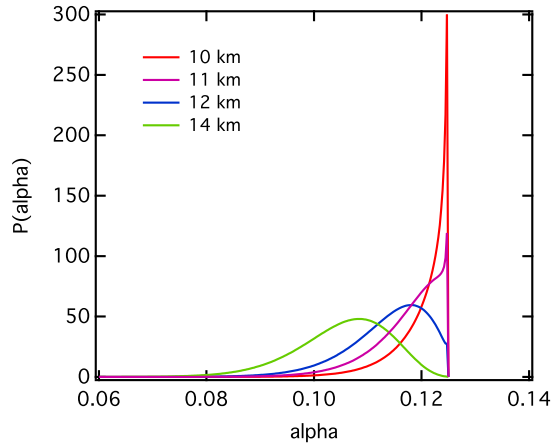}
\caption{{\em (Left)\/} The expected distribution over the parameter
  $\alpha$ calculated for the observed distribution of neutron masses
  and for several values of the neutron star radius that are within the
  physically reasonable range. The possible values of $\alpha$ span an
  extremely narrow range and the likelihood peaks sharply toward the
  critical value $\alpha = 1/8$ for $10-12$~km neutron stars.}
\mbox{}
\label{fig5} 
\end{figure}

We now turn our attention to a second concern arising from the
presence of this limit on $\alpha$. In principle, the situation where
a set of observables with negligible uncertainties yield $\alpha >
1/8$ would imply the inconsistency of the observables and priors with
one another. In practice, however, the two observables $\ftd$ and $A$
have uncertainties and, more importantly, the priors over distance $D$
and the hydrogen mass fraction $X$ are flat over a large range of
values. As a result, the range of inferred $\alpha$ values span a wide
range, with only a fraction of them falling below the $1/8$ limit.
This fraction was used by Steiner et al.\ (2010) to argue for the
inconsistency of observables from some sources; the same argument was
later repeated by Miller (2013). We will now use mock data to explore
this issue quantitatively and will show the shortcomings of this
argument.

We first calculate the expected distribution of $\alpha$ values for
realistic neutron star masses and radii. Using equations~(1), (2), 
and (3), we can write $\alpha$ only in terms of $M$ and $R$ as 
\begin{equation}
\alpha = \frac{GM}{Rc^2} \left(1-\frac{2GM}{Rc^2}\right).
\label{eq:alpha_MR}
\end{equation}
If we assume that all neutron stars have the same radius $R_0$ in the
mass range of interest and masses drawn from the observed distribution
of their descendants (i.e., fast pulsars, see \S2), then we can 
write the expected distribution over $\alpha$ as 
\begin{equation}
P(\alpha) = P(M) \left\vert\frac{dM}{d\alpha}\right\vert.
\label{eq:palpha}
\end{equation}
Equation~(\ref{eq:alpha_MR}) has two solutions
\begin{equation}
M_{\pm} = \frac{Rc^2}{GM} \left(1\pm \sqrt{1-8\alpha}\right)
\end{equation}
and, therefore, for $\alpha\leq 1/8$, 
\begin{equation}
P(\alpha) = \frac{c^2R_0}{G\sqrt{1-8\alpha}}\frac{1}{\sqrt{2\pi\sigma^2_M}}
\left[{\rm e}^{-\frac{(M_- - M_0)^2}{2\sigma^2_M}}+{\rm e}^{-\frac{(M_+ - M_0)^2}{2\sigma^2_M}} \right]
\end{equation}
Here, $M_0= 1.46~M_\odot$ and $\sigma_M = 0.21~M_\odot$ (\"Ozel et
al.\ 2012). We plot in Figure~\ref{fig5} the distribution over
$\alpha$ for several physically reasonable neutron star radii.

For radii in the 10-11~km range, the expected distribution over
$\alpha$ is very narrowly peaked, with an integrable pole at the
limiting value of $\alpha=1/8$. As a result, for such equations of
state, obtaining $\alpha$ values from observations that are close to
the critical value is, in fact, expected, and cannot be taken as an
indication for the inconsistency of the observables, as suggested by
Steiner et al.\ (2010). More importantly, because the two observables
$\ftd$ and $A$ have non-negligible uncertainties and the priors over
$D$ and $X$ are typically flat over a wide range of values, the
observationally inferred distribution of $\alpha$ will be
substantially broader and will very often peak at values $\alpha >
1/8$. It then follows that a very large section of the observationally
inferred distribution over $\alpha$ will be rejected as leading to
unphysical solutions even for favorable uncertainties, without
implying an inconsistency among the observables and the priors.

In order to demonstrate this point, we simulate mock data for a
$1.7~M_\odot$, 10~km neutron star placed at a distance of 5~kpc, with
a hydrogen mass fraction $X=0$ during Eddington-limited bursts and an
atmospheric color correction factor $f_c=1.35$. We assign Gaussian 2\%
uncertainties in the apparent angular size $A$ and in the Eddington
flux $\ftd$; these are comparable to the formal uncertainties reported
by G\"uver et al.\ (2010a) for 4U~1608$-$52. We then draw Monte Carlo
pairs from these two distributions as our mock data for these two
quantities. We also take a mock measured distance that is flat between
$D_{\rm min}$ and $D_{\max}$ such that the true distance lies between
these two values. We also assume, as is sometimes the case, that there
is no measurement of the hydrogen mass fraction and we treat is as a
flat prior between $X=0$ and $X=0.7$.

\begin{figure}
\centering
   \includegraphics[scale=0.45]{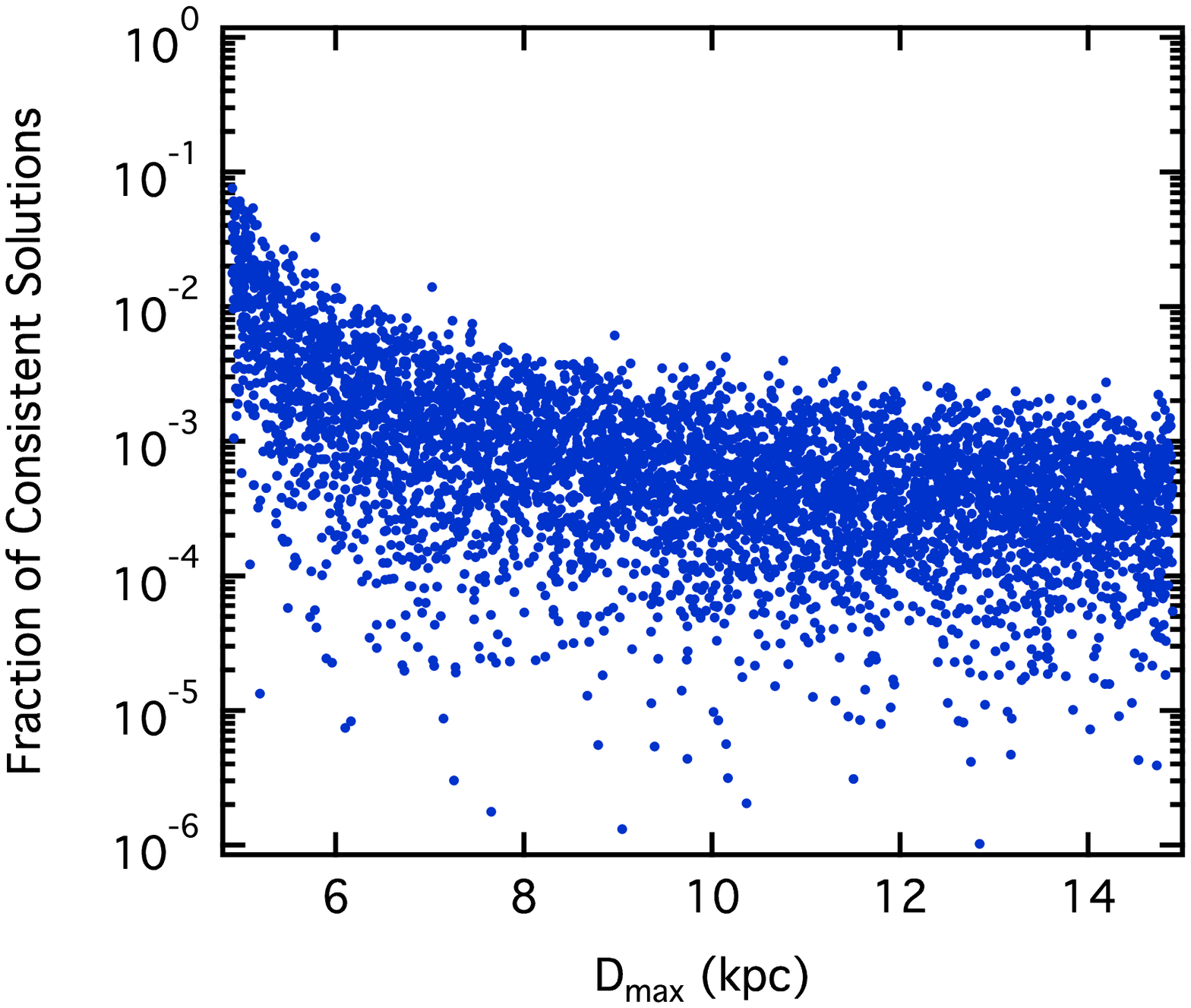}
   \includegraphics[scale=0.45]{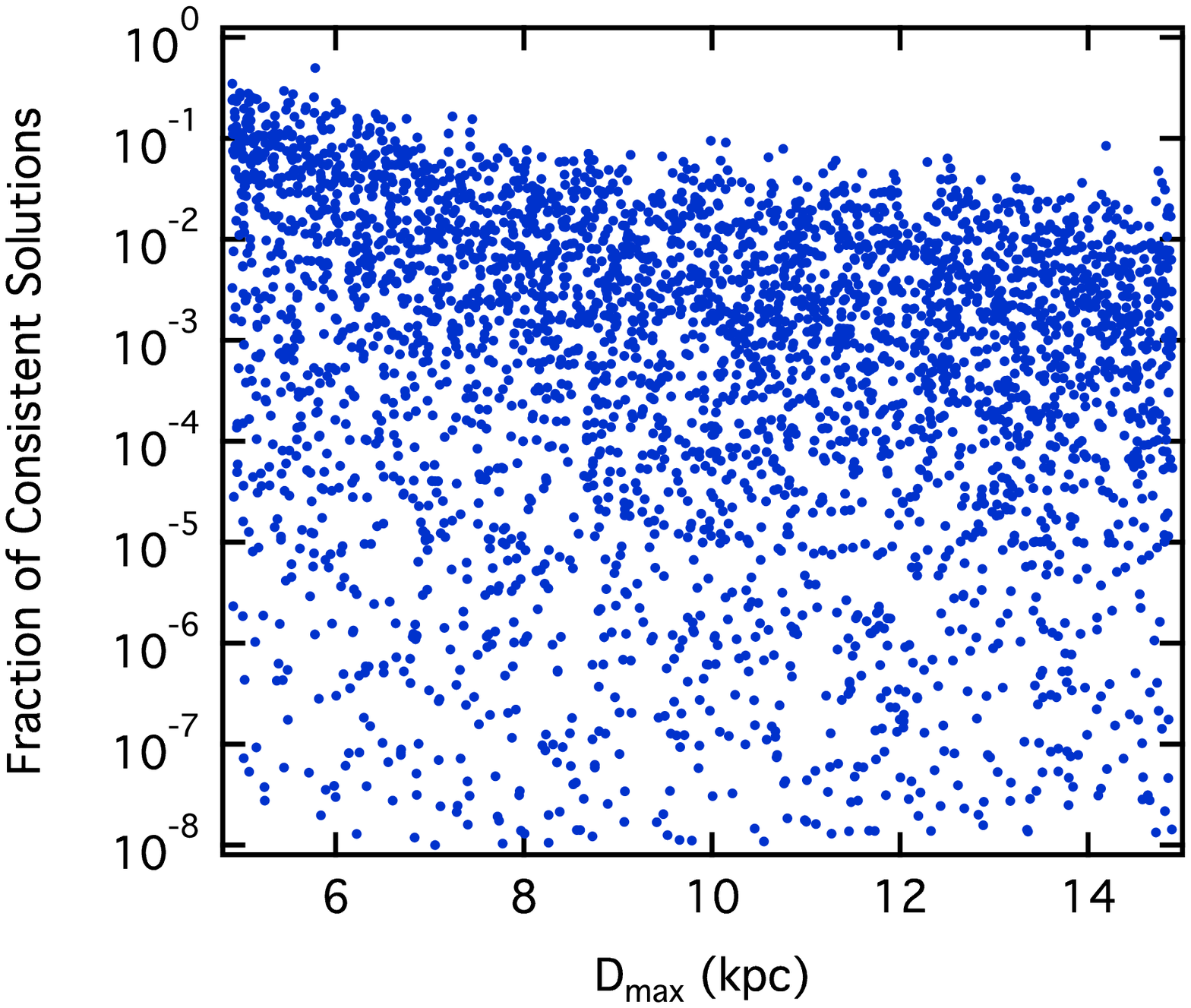}
\caption{The fraction of consistent solutions in the frequentist
  approach as a function of the maximum allowed distance in the
  distance prior. {\it (Top)} The mock data are generated for a
  $1.7~M_\odot$, 10~km neutron star assumed to be at 5~kpc and with
  hydrogen mass fraction $X=0$. They are drawn from Gaussian
  distributions with 2\% uncertainty in in $\ftd$ and $A$, centered at
  the true values for the assumed parameters. In the calculation of
  the fraction of consistent solutions, the priors in the distance and
  the hydrogen mass fraction are assumed to be flat, spanning the
  range from 4.9~kpc to $D_{\rm max}$ and 0.0 to 0.7,
  respectively. The fraction of consistent solutions (see the text for
  definition) is typically $\sim 1\%$ is known accurately a priori and
  is dramatically reduced to $\lesssim 0.1\%$ as the upper bound on
  the source distance increases. {\it (Bottom)} Same as above but with
  true uncertainties in the observables taken to be five times larger
  than the assumed uncertainties. Underestimating the uncertainties by
  a factor of 5 reduces the fraction of consistent solutions to as low
  as $10^{-8}$.}
\label{fig6}
\end{figure}

The fraction of ``accepted'' solutions that correspond to $\alpha \leq
1/8$ can be written as an integral over the entire parameter space, limited 
by the condition
\begin{equation}
\alpha \leq \frac{1}{8} \Rightarrow \ftd\le\ftd^{\rm c}=\frac{c^3 f_c^2 A^{1/2}}{8 k_{\rm 0}(1+X)D}\;.
\end{equation}
such that 
\begin{eqnarray}
\xi & =& \frac{1}{2 \pi \sigma_F \sigma_A (D_{\rm max}-D_{\rm min}) (X_{\rm max}-X_{\rm min})}\nonumber\\ 
&&\quad
\times \int_{D_{\rm min}}^{D_{\rm max}} dD \int_{X_{\rm min}}^{X_{\rm max}} dX \int_0^\infty dA \int_0^{\ftd^c} dF_{\rm td} \nonumber\\
&& \quad  
\times \exp \left[ -\frac{(A - A_{\rm obs})^2}{2\sigma^2_A} - \frac{(\ftd - F_{\rm td, obs})^2}{2\sigma^2_F} \right].
\end{eqnarray}
Here, $A_{\rm obs}$ and $F_{\rm obs}$ are the mock observed data for
each realization and we set the observed uncertainties $\sigma_A$ and
$\sigma_F$ equal to the assumed ones; i.e., the measurements do not
over- or underestimate the true uncertainties. In the top panel of
Figure~\ref{fig6}, we show the fraction $\xi$ of acceptable solutions
as a function of the upper limit that can be placed on the source
distance; for these simulations, we set $D_{\rm min}=4.9$~kpc. Even in
the unrealistic case where the distance is extremely well constrained,
i.e., $D_{\rm max} - D_{\rm min} \le 0.2$~kpc, the fraction of
accepted solutions is equal to at most a few percent. As the upper
bound on the distance increases, the fraction drops to 0.1\% or less.
Such values are comparable to the ones obtained by Steiner et
al.\ (2010) for 1608$-$52, on which these simulations were based, and
show that fractions of this order clearly do not imply any level of
inconsistency.

In the bottom panel of Figure~\ref{fig6}, we show the results of a
similar simulation in which the observational uncertainties are
underestimated by a factor of 5; i.e., the dispersions of the
distributions used to draw the mock data are 10\%, while $\sigma_A$
and $\sigma_F$ are still set to 2\%. We are exploring this situation
because the subsequent comprehensive analysis of all the X-ray data
from sources that show a large number of thermonuclear bursts
indicates a $\simeq 10\%$ spread in the inferred values of $A$ and
$\ftd$ for each source, which is larger than the formal uncertainties
of the individual measurements (G\"uver et al.\ 2012a, b). In this
case, the fraction drops substantially. Steiner et al.\ (2010) used
the similarly low values inferred for 4U~1820$-$30 when only the
statistical uncertainties in the data were considered to justify an ad
hoc and unphysical reinterpretation of the Eddington limit.  This
increased the fraction of consistent solutions by artifically moving
neutron stars to larger radii and thus, away from the critical value
of $\alpha = 1/8$. Instead, a correct assessment of the systematic
uncertainties in the spectroscopic measurements, as was done in the
later studies, provides a much simpler explanation of the inferred
small fraction of accepted solutions, without the need for ad hoc
reinterpretations of the data.

\newpage
\section{Conclusions}

Measuring neutron star radii with spectroscopic and timing techniques
requires at least two observed quantities to break the degeneracies
between the mass and radius introduced by general relativistic
effects. The system of equations that connect the observables to mass
and radius often have critical points and regions of no solutions. In
this paper, we explored the biases in the inferred radii introduced by
this mathematical property of the problem.

We assessed a previously used frequentist method for inferring neutron
star radii, devised a new one within a Bayesian framework, and
compared their performances under frequentist considerations.  We
showed that the former suffers from significant biases in the range of
masses and radii predicted by the modern equations of state and when
realistic uncertainties are taken into account in the measurement of
the distance to the source. In contrast, in the latter framework, the
inferred uncertainties are larger but the most likely values do not
suffer from such biases, making it the preferred framework for
measuring neutron star radii using current and future data.

Finally, we explored ways of quantifying the degree of consistency
between different spectroscopic measurements from a single source.  We
demonstrated that the fraction of the parameter space over the
observables that gives rise to real solutions in mass and radius is
not a good measure of the consistency, as was previously claimed.
Fractions of accepted solutions of order 0.1\% are common due to the
intrinsic uncertainties in the distance and hydrogen mass fraction.
In addition, significantly smaller values of the accepted fraction can
be accounted for by the recently determined systematic uncertainties
in the spectroscopic measurements. In a companion paper, we will apply
the statistical framework discussed here to all the currently
available spectroscopic measurements of the neutron star radii and
constrain the dense matter equation of state.

\acknowledgments

This work was supported by NSF grant AST~1108753. We thank Tolga
G\"uver for useful conversations and comments on the manuscript.  We
thank an anonymous referee for useful suggestions and for helping
improve the presentation of the material.

\end{document}